\documentclass[a4paper,fleqn]{cas-sc}

\usepackage[numbers]{natbib}

\usepackage{algorithm}
\usepackage{algpseudocode}
\usepackage{lineno}

\def\tsc#1{\csdef{#1}{\textsc{\lowercase{#1}}\xspace}}
\tsc{WGM}
\tsc{QE}

\begin{document}
\let\WriteBookmarks\relax
\def\floatpagepagefraction{1}
\def\textpagefraction{.001}

\shorttitle{}    

\shortauthors{X. Xie et~al.}  

\title [mode = title]{Enhanced micrometre spatial resolution and usable photon flux for hybrid pixel detectors via deep-learning-based X-ray localization}



%


\author[a]{X.~Xie} [
    orcid=0000-0001-6473-7886,]
\cormark[1]
\cortext[1]{Corresponding author}
\ead{xiangyu.xie@psi.ch}
\author[a]{A.~Bergamaschi}
\author[a]{M.~Brückner}
\author[a]{M.~Carulla}
\author[a]{T.~Celko}
\author[a]{R.~Dinapoli}
\author[a]{S.~Ebner}
\author[a]{S.~Emiliani}
\author[a]{K.~Ferjaoui}
\author[a]{E.~Fröjdh}
\author[a]{V.~Gautam}
\author[a]{D.~Greiffenberg}
\author[a]{S.~Hasanaj}
\author[a]{V.~Hinger}
\author[a]{M.~Hürst}
\author[a]{V.~Kedych}
\author[a]{T.~King}
\author[a]{S.~Li}
\author[a]{C.~Lopez-Cuenca}
\author[a]{L.~Lunin}
\author[a]{F.~Marone}
\author[a]{A.~Mazzoleni}
\author[a]{D.~Mezza}
\author[a]{S.~Moses}
\author[a]{K.~Moustakas}
\author[a]{A.~Mozzanica}
\author[a]{J.~Mulvey}
\author[a]{M.~Müller}
\author[a]{C.~Ruder}
\author[a]{B.~Schmitt}
\author[a]{S.~Silletta}
\author[a]{D.~Thattil}
\author[a]{J.~Zhang}




\affiliation[a]{organization={PSI Center for Photon Science, Paul Scherrer Institute},
            addressline={Forschungsstrasse 111}, 
            city={Villigen},
            postcode={5232}, 
            country={Switzerland}}




\begin{abstract}
    Micrometre spatial resolution is achieved for X-rays with the M\"ONCH detector, a charge-integrating hybrid pixel detector with a 25~\textmu m pixel pitch, by exploiting charge sharing between neighbouring pixels.
    The conventional analytical interpolation method, however, relies on a global mapping derived from the charge-sharing statistics and is applicable only to isolated single-photon clusters, which limits both the achievable resolution and the usable photon flux.
    To overcome these limitations, we trained deep learning models to localize single-photon and pile-up events based on high-fidelity simulation data.

    For knife-edge measurements at 12 keV and a low occupancy of 0.74\% (photons per pixel and frame), where single-photon events dominate, the proposed method reaches a $\sigma$ of the edge spread function of 1.36~\textmu m, compared with 1.54~\textmu m for the conventional eta interpolation.
    We demonstrate experimentally that resolving up to three photons per cluster raises the reconstruction efficiency from 59\% to 93\% at a $2.4\times$ higher occupancy of 1.81\%, while preserving a $\sigma$ of 1.39~\textmu m.
\end{abstract}





\maketitle


\newcommand{\phfig}[2]{\fbox{\parbox[c][#1][c]{0.95\linewidth}{\centering\itshape #2}}}

\section{Introduction}
\label{sec:introduction}

M\"{O}NCH, a charge-integrating hybrid pixel detector with a 25~\textmu m pixel pitch \cite{Ramilli_2017}, provides a unique combination of high spatial resolution, kHz frame rate, and energy discrimination for X-ray imaging \cite{Cartier_2014, Bergamaschi02112018}.
The small pixel size enables micrometre spatial resolution for X-rays \cite{Cartier:yn5011, Chiriotti_2022, Qu:26} by exploiting charge sharing via the eta interpolation method \cite{BELAU1983253}: the charge-weighted centre $\eta$ of the single-photon cluster is mapped to the sub-pixel position using a single global mapping derived from the charge-sharing statistics.
This approach, however, has two limitations regarding the achievable spatial resolution and the usable photon flux.
Given the nature of X-ray absorption and the resulting variation in charge sharing with the photon absorption depth, a single global mapping cannot accurately describe the sub-pixel position for all events, which limits the achievable spatial resolution.
Since it requires isolated single-photon clusters, the eta interpolation is applicable only to low-flux measurements, which limits the usable photon flux and thus the experimental throughput.

\begin{figure}
  \centering
  \includegraphics[width=0.48\linewidth]{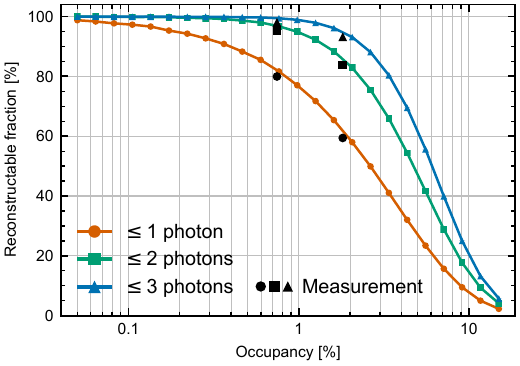}
  \caption{Fraction of the reconstructable photons as a function of
  the detector occupancy from simulation and measurement, when clusters containing at most one, two, or three photons are accepted. 
  Frames with Poisson-distributed 12~keV photons were processed with the cluster finder and energy windows of Section~\ref{sec:detector}.}
  \label{fig:occupancy}
\end{figure}

Diffraction-limited storage rings, such as MAX~IV \cite{Tavares:xe5003}, ESRF-EBS \cite{Raimondi2023} and SLS~2.0 \cite{Streun2018}, provide up to two orders of magnitude higher brilliance.
Applications which require a detector with micrometre spatial resolution together with a high photon flux, such as X-ray Fourier ptychography \cite{doi:10.1126/sciadv.aav0282}, can therefore be carried out with a substantially higher experimental throughput.
Meanwhile, the micrometre detector resolution of M\"ONCH also benefits lab-based X-ray imaging, e.g.\ full-field transmission X-ray microscopy with sub-micrometre resolution \cite{Qu:26, Qu:26PR} and beam-tracking multimodal imaging \cite{Dreier_2020}.
Under these conditions the requirement of single-photon clusters limits the photon flux and results in long measurement times.
The fraction of photons which can be reconstructed decreases rapidly with the detector occupancy, defined here as the average number of photons per pixel and frame\footnote{Owing to charge sharing, the charge of a single photon can spread over up to $3 \times 3$ pixels.}, as shown in Fig.~\ref{fig:occupancy}.

At an occupancy of 1.81\%, corresponding to the high-flux measurement of Section~\ref{sec:exp-highflux}, only 60\% of the photons are expected to be contained in single-photon clusters, while the remaining photons have to be discarded.
Pile-up events rapidly dominate at higher occupancies.
Increasing the frame rate to reduce the occupancy poses significant challenges for the readout electronics and the data acquisition system.
On the other hand, resolving pile-up events at the data analysis level can significantly increase the usable photon flux without any hardware modification.

Deep learning (DL) provides a natural way to overcome both limitations of the eta interpolation, namely the resolution loss caused by the single global mapping and the flux restriction imposed by the single-photon requirement.
Rather than reducing a cluster to its charge-weighted centre, a neural network learns the nonlinear relation between the full charge-sharing pattern sampled by the pixel array and the photon position.
It can therefore adapt to the charge cloud of each individual event, and it can be trained to localize several photons within the same pile-up cluster.
Neural networks have been applied to position reconstruction in pixelated detectors for charged particles \cite{ATLAS_2014}, electrons \cite{VANSCHAYCK2020113091} and photons \cite{Yang_2023, Peirson_2021}.
For the M\"ONCH detector, DL-based localization has been demonstrated to improve the position resolution for 200~keV electrons in electron microscopy \cite{Xie_2024_EM} with respect to analytical methods.

The transfer of the DL approach to X-rays has so far been hindered by the lack of suitable training samples, since measured single-photon events whose incident positions are known with micrometre precision are difficult to obtain, let alone pile-up events.
Two recent developments remove this obstacle.
First, a charge transport simulation for X-rays incorporating the repulsion effect \cite{XIE2026170894} provides high-fidelity simulated events with the ground truth.
Second, a pixel-wise gain and nonlinearity calibration \cite{XIE2026171227} improves both the energy accuracy and the uniformity of the pixel response, reducing the systematic differences between simulation and measurements.

In this work, deep learning models were developed to localize single-photon and pile-up events in the M\"ONCH detector, trained on simulated events.
The models were validated with knife-edge measurements at 12~keV at the DanMAX beamline of MAX~IV.
For single-photon events measured at the 0.74\% low-flux condition, the edge spread function width $\sigma$ is improved to $1.36\pm0.02$~\textmu m with respect to $1.54\pm0.02$~\textmu m of the eta interpolation.
By resolving clusters containing up to three photons, the reconstruction efficiency is increased from 59\% to 93\% at an occupancy of 1.81\% while preserving a $\sigma$ of $1.39\pm0.02$~\textmu m.
This paper is organized as follows.
Section~\ref{sec:reconstruction} describes the detector, the reconstruction framework, the generation of the training samples and the network architecture.
Section~\ref{sec:simulation} presents the localization performance obtained in simulation for single-photon and pile-up events, and Section~\ref{sec:experiment} the experimental validation.
Section~\ref{sec:discussion} discusses the origin of the improvement, the quantitative comparison between simulation and measurement, the necessity of the pile-up reconstruction, the limitations and the computational cost, followed by the conclusions in Section~\ref{sec:conclusion}.

\section{Detector and reconstruction framework}
\label{sec:reconstruction}

\subsection{The M\"ONCH03 detector and data processing pipeline}
\label{sec:detector}

The M\"ONCH03 detector \cite{Ramilli_2017} used in this experiment consists of a $400 \times 400$ pixel array with a pixel pitch of 25~\textmu m, bump-bonded to a 650~\textmu m-thick silicon sensor.
Each pixel comprises a charge-sensitive amplifier, a correlated double sampler and configurable gains, and the analog outputs of every $25 \times 200$ pixel block are digitized by one 14-bit ADC channel.
The fully analog, charge-integrating readout preserves the charge shared among neighbouring pixels.
For the measurements reported here the detector was operated in the high gain mode at an exposure time of 50~\textmu s, with the silicon sensor biased at 150~V and a cooling temperature of 20~$^{\circ}$C, corresponding to a noise of 0.13~keV RMS per pixel.

\begin{figure}
  \centering
  \includegraphics[width=0.85\linewidth]{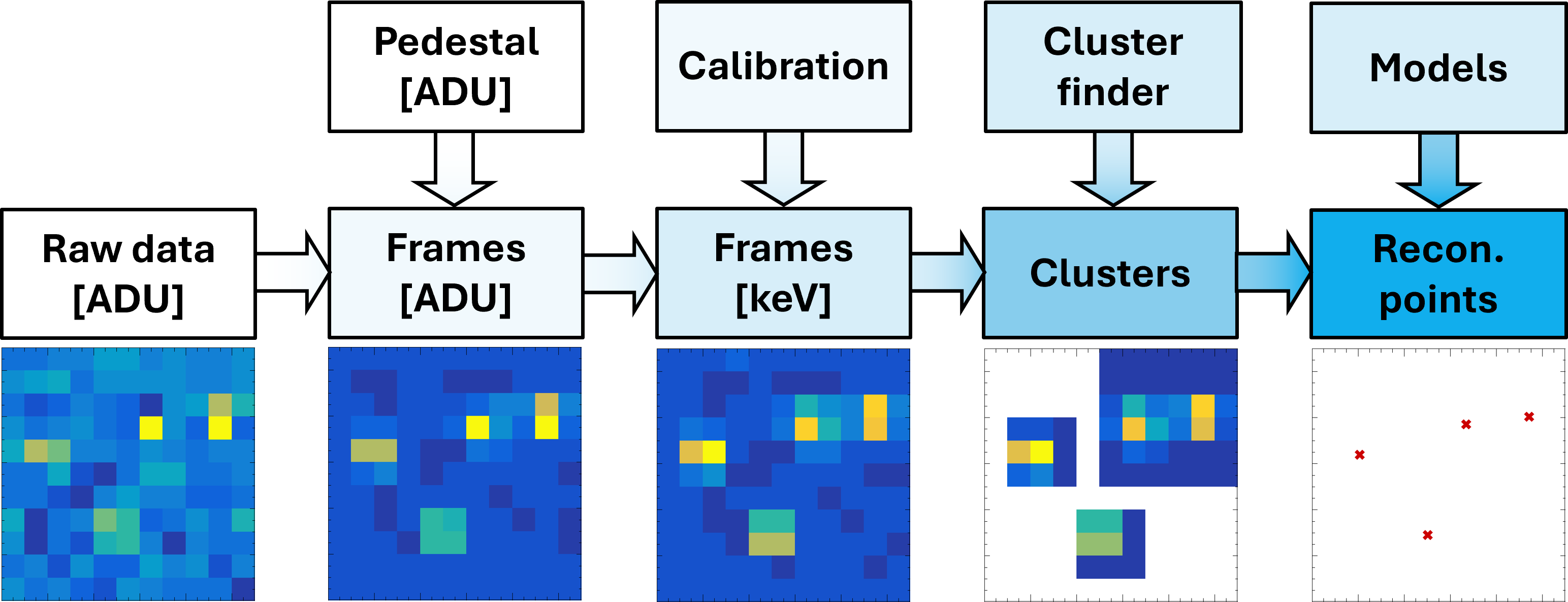}
  \caption{Data processing pipeline. Raw frames are pedestal-corrected and
  converted to energy using a pixel-wise lookup table; clusters are then extracted and passed to the deep learning models and the eta interpolation (for single-photon events), which return the incident positions of the photons in the
  cluster. The reconstructed sub-pixel positions are flattened before the image is
  accumulated, as is the case for the eta interpolation.}
  \label{fig:pipeline}
\end{figure}

The data processing pipeline is shown in Fig.~\ref{fig:pipeline}.
Raw frames in analog-to-digital units (ADU) are first corrected for the pedestal, and are subsequently converted to energy by means of the pixel-wise lookup table described in \cite{XIE2026171227}.
Clusters are then formed by selecting a seed pixel with a signal exceeding five times its noise level and recursively adding neighbouring pixels with a signal above three times their noise level \cite{Fröjdh_2025}.
The multiplicity $N$ of a cluster, i.e.\ the number of photons it contains, is determined from its total energy: a cluster is assigned multiplicity $N$ if its energy lies within $N \times (12 \pm 2)$~keV and clusters with $N \le 3$ are considered in this work.
For each cluster, a window of $3N \times 3N$ pixels ($3 \times 3$, $6 \times 6$ and $9 \times 9$ for $N = 1, 2, 3$) centred on its charge-weighted centre is extracted.
Pixels in the window which do not belong to the cluster are retained, since they may still contain charge from the photon(s) of interest below the clustering threshold, unless they are assigned to another cluster, in which case they are set to zero.
Clusters are passed to the deep learning models to reconstruct the incident position(s) of the photon(s) in the cluster.
The sub-pixel positions are subsequently flattened to avoid grid artefacts as described in Section~\ref{sec:eta}.
Eta interpolation \cite{BELAU1983253} is used as the reference method for single-photon events.

\subsection{Training samples generated from simulation}
\label{sec:training-samples}

The 12~keV X-ray samples were generated using the charge transport simulation incorporating the repulsion effect described in \cite{XIE2026170894}.
Charge carrier-wise simulations were carried out at 150~V bias and 307.8~K sensor temperature (measured at 20~$^{\circ}$C cooling temperature) for 128 absorption depths, uniformly distributed over the 650~\textmu m sensor thickness.
For each event, the photon absorption depth was sampled from an exponential distribution and the corresponding drift time was calculated.
The distribution of the collected charge carriers was then sampled from the generalized Gaussian distribution parameterized as a function of the drift time.
A Fano factor of 0.13 was applied in order to model the statistical fluctuations of the number of charge carriers.
Random sub-pixel shifts were introduced in the $x$ and $y$ coordinates to ensure the generalization of the network.
A Gaussian noise of 0.13~keV RMS was added to each pixel.
A total of 10~M single-photon events were generated in a $3 \times 3$ pixel window.

Pile-up samples were obtained by superimposing two or three random single-photon events within one cluster window, with the positions uniformly distributed over the window of $6 \times 6$ and $9 \times 9$ pixels respectively.
The same Gaussian noise of 0.13~keV RMS was added to each pixel after the superposition.
To ensure the consistency with the experimental data, the same clustering algorithm was applied to the superimposed events, and those events with more than one cluster were discarded.
16~M pile-up events were generated for each of the two-photon and three-photon samples.
The datasets were split into 81\% for training, 13\% for validation and 6\% for testing.

\subsection{Neural network architecture and training}
\label{sec:architecture}

The network input consists of the calibrated pixel energies of the cluster and the $x$ and $y$ coordinates embedded in the cluster window \cite{liu2018coordconv}, three channels in total.
Given the translation equivariance of the convolutional layers, the coordinate embedding facilitates the learning of the absolute position within the cluster window.
One network was trained for each cluster multiplicity.

Each network consists of a stack of two-dimensional convolutional layers, which extract features of the charge-sharing pattern.
The following fully connected layers return the $2N$ coordinates of the $N$ photons in the cluster.
In the two-photon and three-photon networks, a spatial attention module \cite{woo2018cbam} weights each position of the feature map before the fully connected layers.
Since only part of the cluster window contains the charge belonging to a given photon, this allows the contributions of the individual charge clouds to be weighted separately.
The networks are compact, with 23.7~k trainable parameters for the single-photon network and 1.21~M and 1.95~M for the two-photon and three-photon networks, respectively.
The detailed layer configuration is available in the repository given in the data availability statement.


The models were trained by minimizing a smooth L1 loss with $\beta = 0.01$ in units of the pixel pitch. 
For clusters containing more than one photon the network returns a set of positions without a defined order, and the loss must therefore be invariant under permutations of the predictions.
The loss is evaluated for all assignments between the $N$ predicted and the $N$ true positions, and the smallest value is used for back-propagation.
The training was performed with PyTorch \cite{PyTorch} using the Adam optimizer with an initial learning rate of 0.001 for 500 epochs on an NVIDIA RTX 4090 GPU, with a training time up to 12~h.
The trained weights and the complete training configuration are available in the same repository.

\subsection{Eta interpolation and flattening of the sub-pixel distribution}
\label{sec:eta}

In the eta interpolation as the reference method, the charge-weighted centre $\eta_x$ and $\eta_y$ are first calculated from the $3 \times 3$ cluster.
Since the resulting two-dimensional eta distribution is non-uniform, it is flattened by means of a Rosenblatt transform \cite{rosenblatt1952remarks}, which maps it onto a uniform sub-pixel distribution.
The interpolated sub-pixel position is then obtained by applying bilinear interpolation using lookup tables derived from the transform.

\begin{figure}
  \centering
  \includegraphics[width=0.48\linewidth]{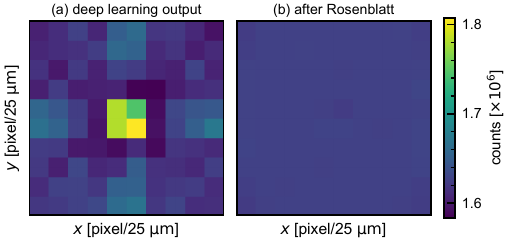}
  \caption{Sub-pixel distribution of the reconstructed positions for flat-field illuminated measurement samples, (a) as returned by the network and (b) after the Rosenblatt transform. The excess at the pixel centre and along the two axes through it in (a) originates from the events whose charge cloud is confined to one or two pixels, thus no sub-pixel information in both or in one of the two coordinates is available.}
  \label{fig:subpixel}
\end{figure}

The same Rosenblatt transform step is required for the output of the networks, for a reason which originates in the physics of the charge deposition.
When the charge cloud is confined to one or two pixels due to deep absorption, the sub-pixel position is not measurable for two or one of the coordinates, respectively.
By minimizing the residual, the network returns the pixel centre for these coordinates, which leads to a bright spot at the pixel centre and to a cross-shaped pattern in the sub-pixel distribution of the reconstructed positions, as illustrated in Fig.~\ref{fig:subpixel}(a).
The Rosenblatt transform is derived for each dataset and applied to the network output, which restores a uniform distribution, Fig.~\ref{fig:subpixel}(b), at the cost of a negligible increase of the width of residual distribution (less than 0.01~\textmu m in simulation).
All results reported in the following were obtained after flattening, for both methods.

\section{Simulation results}
\label{sec:simulation}

\subsection{Single-photon localization}
\label{sec:sim-single}

\begin{figure}
  \centering
  \includegraphics[width=0.48\linewidth]{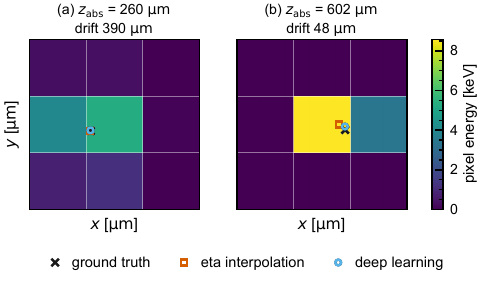}
  \caption{Typical simulated single-photon events for a shallow (a) and a deep (b) absorption depth, showing the ground truth position together with the positions reconstructed by the eta interpolation and by the network. The deep-absorption event produces a compact charge cloud for which the eta mapping is systematically biased.}
  \label{fig:sim-events-single}
\end{figure}

Two typical simulated single-photon events are shown in Fig.~\ref{fig:sim-events-single}, together with the ground truth and reconstructed positions.
For event (a) with a shallow absorption and a larger cloud size, both methods reconstruct the position accurately.
For event (b) with a deep absorption and a smaller cloud size, the position reconstructed by the eta interpolation is biased towards the pixel centre while the network returns a more accurate position.

\begin{figure}
  \centering
  \includegraphics[width=0.48\linewidth]{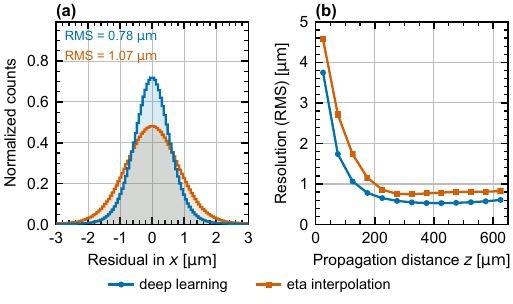}
  \caption{(a) Distribution of the residual between the reconstructed and the true incident position for the eta interpolation and for the network, for 12~keV photons. (b) Spatial resolution, given as the RMS of the residual, as a function of the charge propagation distance.}
  \label{fig:resolution}
\end{figure}

Fig.~\ref{fig:resolution}(a) shows the residual distributions obtained at 12~keV.
Throughout this work the spatial resolution in simulation is quoted as the RMS of the one-dimensional residual between the reconstructed and the true incident position.
An RMS of 0.78~\textmu m is obtained with the deep learning model, a 27\% improvement compared with 1.07~\textmu m for the eta interpolation.
The dependence of the spatial resolution on the charge propagation distance, i.e.\ the distance between the absorption point and the readout electrode, is shown in Fig.~\ref{fig:resolution}(b).
The spatial resolution degrades rapidly for propagation distances below approximately 250~\textmu m.
Eta interpolation shows a minimum at 250~\textmu m, which is further discussed in Section~\ref{sec:discussion}.
For longer propagation distances the charge cloud is large enough to cover several pixels, and the resolution converges to 0.6~\textmu m and 0.8~\textmu m for the deep learning model and the eta interpolation, respectively.
The deep learning model outperforms the eta interpolation over the full range, and the absolute improvement is largest at short propagation distances, where the assumption of a single global mapping exhibits the largest deviation.
Since the photons enter the sensor from the side opposite to the pixel electrodes, only about 11\% of the absorbed photons have a propagation distance below 50~\textmu m\footnote{The absorption depth of the X-rays follows an exponential distribution, with an attenuation length of about 220~\textmu m in silicon at 12~keV.}.
The overall resolution in Fig.~\ref{fig:resolution}(a) is therefore dominated by the long-propagation regime, while the region of largest improvement gains weight at higher photon energies (Section~\ref{sec:discussion}).

\subsection{Pile-up localization}
\label{sec:sim-pileup}

\begin{figure}
  \centering
  \includegraphics[width=0.48\linewidth]{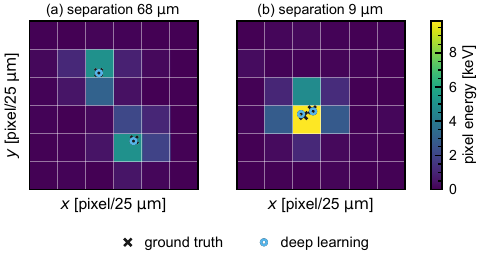}

  \includegraphics[width=0.48\linewidth]{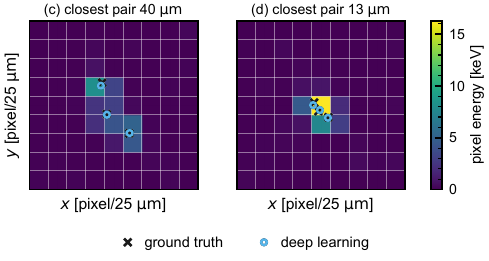}
  \caption{Typical simulated pile-up events with two (a, b) and three (c, d)
  photons in one cluster, showing the ground truth positions and the positions
  predicted by the corresponding networks.}
  \label{fig:sim-events-pileup}
\end{figure}

Typical simulated pile-up events containing two and three photons are shown in Fig.~\ref{fig:sim-events-pileup}, together with the ground truth positions and the predictions of the corresponding networks.
Conventional clustering methods can no longer separate the photons with partially or strongly overlapping charge clouds, while the networks are able to reconstruct the positions of all photons in the cluster accurately.

\begin{figure}
  \centering
  \includegraphics[width=0.95\linewidth]{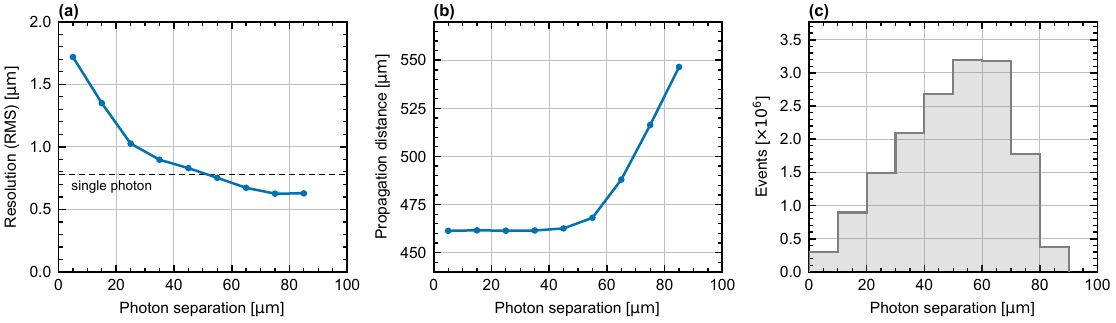}
  \caption{
    (a) Resolution as the RMS of the residuals for the two-photon predictions as a function of the distance between the two photons in the cluster. 
    (b) Average propagation distance of the two photons as a function of their separation.
    (c) Distribution of the two-photon pile-up events categorized by their separation distance.
    The resolution improves with increasing separation, and a better resolution than the single-photon case is obtained for separations above 50~\textmu m, which is attributed to the larger charge clouds required by the larger separation to form pile-up events.
  }
  \label{fig:sim-resolution-separation}
\end{figure}

The impact of the pile-up on the achievable spatial resolution is investigated using the two-photon pile-up events.
Fig.~\ref{fig:sim-resolution-separation}(a) shows the RMS of the residuals as a function of the distance between the two photons in the cluster:
worse resolution is observed for smaller separations, as expected from the increasing overlap of the charge clouds.
The resolution gradually improves with increasing separation, and a better resolution than the single-photon case is obtained for separations above 50~\textmu m.
This is attributed to the larger charge clouds required by the larger separation to form pile-up events, as shown in Fig.~\ref{fig:sim-resolution-separation}(b): these events with larger separations need large enough charge clouds to form a single cluster, which carry more sub-pixel information and thus allow for a better localization.
The saturated resolution at large separations is 0.6~\textmu m, consistent with that of single-photon events with long propagation distances (Fig.~\ref{fig:resolution}(b)).
Since the RMS is dominated by the events with small separations, where the residuals are considerably larger, the overall two-photon resolution of 0.86~\textmu m nevertheless remains slightly worse than the single-photon one.

\begin{table}
  \centering
  \caption{Summary of the spatial resolution obtained in simulation for single-photon and pile-up events at 12~keV. The resolution is given as the RMS of the residuals between the reconstructed and the true incident positions.}
  \label{tab:sim-summary}
  \begin{tabular}{lcc}
    \hline
    Cluster multiplicity & Method & Resolution (RMS) \\
    \hline
    Single-photon & eta & 1.07~\textmu m \\
    Single-photon & DL & 0.78~\textmu m \\
    Two-photon & DL & 0.86~\textmu m \\
    Three-photon & DL & 1.09~\textmu m \\
    \hline
  \end{tabular}
\end{table}

The results of all configurations are summarized in Table~\ref{tab:sim-summary}.
The resolution achieved via deep learning degrades moderately from 0.78~\textmu m to 0.86~\textmu m and 1.09~\textmu m for clusters containing one, two and three photons, respectively.
Compared with the eta interpolation, however, the deep learning model exhibits a 27\% improvement in spatial resolution for single-photon events and the capability to resolve pile-up events with a comparable resolution.

\section{Experimental validation}
\label{sec:experiment}

\subsection{Measurement setup and analysis procedure}
\label{sec:exp-setup}
To experimentally validate the deep learning method, knife-edge measurements were carried out at the DanMAX beamline of the MAX~IV Laboratory.
A monochromatic X-ray beam at a photon energy of 12~keV was used, with two flux levels obtained by inserting a different number of aluminium filters of 0.22~mm thickness each. 
The beam size of 1.6~mm in height and 1.3~mm in width provided flat-field illumination over the region of interest. 
A polished tungsten knife-edge mask was mounted a few millimetres upstream of the sensor and was slightly tilted by 4.9~$^{\circ}$ with respect to the pixel columns.
Flat-field images were acquired without the mask to correct the non-uniformity in the beam profile.
14~M and 7~M frames were acquired for the low- and high-flux conditions, respectively, ensuring around $10^{5}$~photons per pixel in the illuminated region.

The occupancies of the low- and high-flux conditions were determined from the flat-field images over the region of interest used for the edge spread function analysis (336 pixels), as the total energy of the clusters divided by the mean energy of a single-photon cluster, the number of pixels and the number of frames, yielding 0.74\% and 1.81\%, respectively.
The reconstruction efficiency, i.e.\ the fraction of the absorbed photons which are reconstructed, was estimated as the ratio of the summed energy contained in the reconstructable clusters to the total energy of the same region of interest.

\begin{figure}
  \centering
  \includegraphics[width=0.48\linewidth]{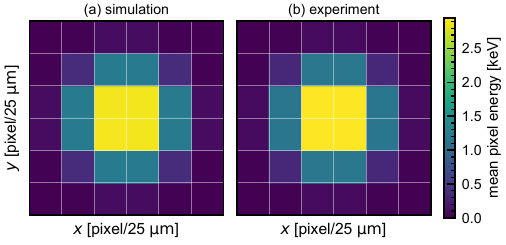}
  \caption{Average two-photon clusters from the simulation and measurement to check the consistency of the charge transport simulation and the clustering algorithm.
  The projected 1D RMS of the two-photon cluster is 0.955~pixels and 0.934~pixels for the simulation and measurement, respectively.
  }
  \label{fig:2ph_average_sample}
\end{figure}
The data were processed with the pipeline shown in Fig.~\ref{fig:pipeline}.
The average two-photon sample from simulation and measurement is shown in Fig.~\ref{fig:2ph_average_sample}, validating the consistency of the processing pipelines and the charge transport simulation.
The sub-pixel reconstructed positions were flattened via the Rosenblatt transform.
All images were reconstructed in bins of 2.5~\textmu m (one tenth of the pixel pitch) and were corrected with the subsequent flat-field measurement.
The subdivision of each pixel into ten bins follows previous works with M\"ONCH \cite{Chiriotti_2022} and balances the acquisition time against the statistics per bin, which decreases with the bin area.

The spatial resolution was evaluated from the edge spread function (ESF), which was obtained by projecting the reconstructed intensity onto the axis perpendicular to the edge.
The ESF was fitted with an error function, from which the width $\sigma$ was extracted.
The binning adds a contribution of $2.5/\sqrt{12} \approx 0.72$~\textmu m in quadrature to $\sigma$, i.e.\ the RMS of a uniform distribution over one bin, which is included in all reported widths unless stated otherwise.
The modulation transfer function (MTF) was calculated via the Fourier transform of the discrete line spread function (LSF), obtained by differentiating the ESF.
The spatial frequency at which the MTF drops to 50\%, denoted MTF50, is reported\footnote{The more commonly used MTF10 already exceeds the Nyquist frequency of 200~lp/mm for the 2.5~\textmu m sub-pixel binning, and is therefore not reported. The reported MTF values include the aperture of the 2.5~\textmu m bins, which alone reduces the MTF to $2/\pi \approx 0.64$ at the Nyquist frequency.} together with the MTF at the Nyquist frequency of 200~lp/mm.
\subsection{Low flux: the single-photon dominated regime}
\label{sec:exp-lowflux}
\begin{figure}
  \centering
  \includegraphics[width=0.70\linewidth]{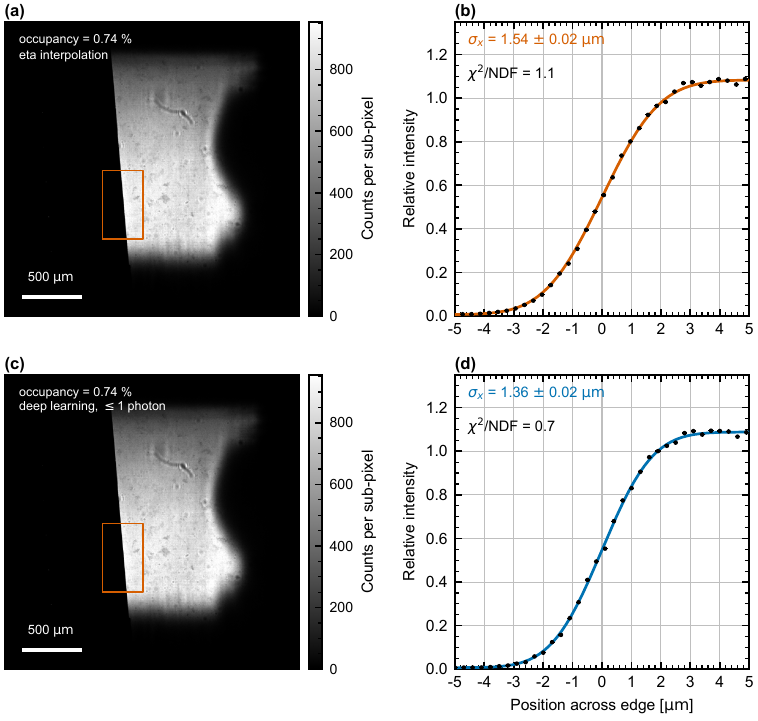}
  \caption{
    Knife-edge measurement at 12~keV and 0.74\% occupancy, reconstructed accepting only single-photon clusters via (a) the eta interpolation and (c) the deep learning model with the region of interest indicated by the orange box.
    (b, d) plot the corresponding edge spread functions and error-function fits.
  }
  \label{fig:exp-lowflux}
\end{figure}

At the lower occupancy of 0.74\%, 80\% of the photons are contained in single-photon clusters, where the eta interpolation and deep learning can be directly compared.
The knife-edge image is shown in Fig.~\ref{fig:exp-lowflux} together with the corresponding ESF fits on the flat-field-corrected images.
A width of $\sigma = 1.36 \pm 0.02$~\textmu m is obtained with the deep learning model, a 12\% improvement compared with $\sigma = 1.54 \pm 0.02$~\textmu m for the eta interpolation.
After subtracting the binning contribution in quadrature, the intrinsic widths correspond to 1.15~\textmu m and 1.36~\textmu m, respectively, i.e.\ a 15\% improvement.
The corresponding MTF50 values are 141~lp/mm and 127~lp/mm.
It should be noted that the relative intensity on the illuminated side is higher than 1, indicating an excess of single-photon events close to the edge: photons landing next to the shadowed region have fewer neighbours and are therefore less likely to form a pile-up cluster.
This effect and the difference between the measured widths and the widths expected from simulation are discussed in Section~\ref{sec:discussion}.

In this low-flux case, accepting clusters with up to two and three photons already raises the reconstruction efficiency to 95\% and 98\%, respectively, while the width of the ESF remains at $1.38 \pm 0.02$~\textmu m in both cases.

\subsection{High flux: the pile-up regime}
\label{sec:exp-highflux}

\begin{figure}
  \centering
  \includegraphics[width=0.70\linewidth]{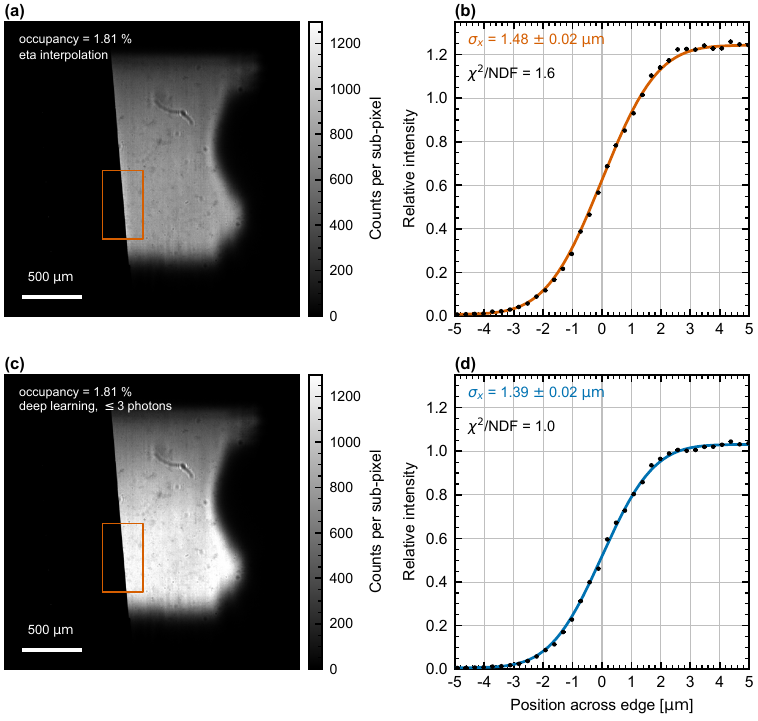}
  \caption{
    Knife-edge measurement at 12~keV and 1.81\% occupancy, reconstructed accepting only single-photon clusters via (a) the eta interpolation and (c) accepting up to three-photon clusters via the deep learning model with the region of interest indicated by the orange box.
    (b, d) plot the corresponding edge spread functions and error-function fits.
  }
  \label{fig:exp-highflux}
\end{figure}

At the higher occupancy of 1.81\%, only 59\% of the photons are contained in single-photon clusters.
The images reconstructed with the eta interpolation and with the deep learning models accepting clusters with up to three photons are shown in Fig.~\ref{fig:exp-highflux}, together with the corresponding ESF fits.
When only single-photon clusters are accepted, a bright rim is observed along the edge, since photons next to the shadowed region have fewer neighbours and are therefore less likely to be rejected as pile-up.
The resulting intensity-dependent reconstruction efficiency, discussed in Section~\ref{sec:discussion}, raises the relative intensity next to the edge to 1.24 and leads to a worse ESF fitting quality.
With the eta interpolation, $\sigma = 1.48 \pm 0.02$~\textmu m ($\chi^2/\mathrm{NDF} = 1.6$) and an MTF50 of 132~lp/mm are obtained, compared with $1.54 \pm 0.02$~\textmu m and 127~lp/mm at low flux.
The difference is mainly attributed to the worse fitting quality, and the two results are compatible within 2 standard deviations.
Applying the deep learning model to the same single-photon clusters gives $\sigma = 1.35 \pm 0.02$~\textmu m, consistent with the low-flux result, and an MTF50 of 145~lp/mm.

Accepting clusters with up to three photons increases the reconstruction efficiency to 93\%.
The width of the edge spread function remains at $1.39 \pm 0.02$~\textmu m with a corresponding MTF50 value of 140~lp/mm.
Compared with the low-flux case, the resolution remains unchanged while extending the photon flux by a factor of 2.4 and removing the excess intensity next to the edge.
Reconstruction efficiency, resolution and MTF50 values are summarized in Table~\ref{tab:exp-summary} for the two occupancies and for the two reconstruction methods.

\begin{table}[t]
  \centering
  \caption{Summary of the knife-edge measurements at 12~keV: reconstruction efficiency, width $\sigma$ of the edge spread function, MTF50 (the spatial frequency at 50\% contrast) and the MTF at the Nyquist frequency of 200~lp/mm, for the two occupancies and for the different reconstruction
  methods. DL denotes the deep learning models. The ESF widths include the binning contribution of 0.72~\textmu m.}
  \label{tab:exp-summary}
  \begin{tabular}{lcccccc}
    \hline
    Occupancy & Method & \begin{tabular}[b]{@{}c@{}}Accepted\\ multiplicity\end{tabular} & \begin{tabular}[b]{@{}c@{}}Reconstruction\\ efficiency\end{tabular}  & ESF width ($\sigma$) & MTF50 & MTF(200~lp/mm) \\
    \hline
    0.74\% & eta & $\leq$1 photon  & 80\% & $1.54 \pm 0.02$~\textmu m & 127~lp/mm & 0.170 \\
    0.74\% & DL  & $\leq$1 photon  & 80\% & $1.36 \pm 0.02$~\textmu m & 141~lp/mm & 0.225 \\
    0.74\% & DL  & $\leq$2 photons & 95\% & $1.38 \pm 0.02$~\textmu m & 138~lp/mm & 0.221 \\
    0.74\% & DL  & $\leq$3 photons & 98\% & $1.38 \pm 0.02$~\textmu m & 137~lp/mm & 0.229 \\
    \hline
    1.81\% & eta & $\leq$1 photon  & 59\% & $1.48 \pm 0.02$~\textmu m & 132~lp/mm & 0.187 \\
    1.81\% & DL  & $\leq$1 photon  & 59\% & $1.35 \pm 0.02$~\textmu m & 145~lp/mm & 0.229 \\
    1.81\% & DL  & $\leq$2 photons & 84\% & $1.37 \pm 0.02$~\textmu m & 143~lp/mm & 0.223 \\
    1.81\% & DL  & $\leq$3 photons & 93\% & $1.39 \pm 0.02$~\textmu m & 140~lp/mm & 0.244 \\

    \hline
  \end{tabular}

\end{table}

\section{Discussion}
\label{sec:discussion}

\textbf{Origin of the improved spatial resolution}: The eta interpolation applies a single mapping to all clusters, irrespective of the absorption depth and thus of the size of the charge cloud.
The optimal resolution is achieved at a propagation distance of about 250~\textmu m (Fig.~\ref{fig:resolution}(b)), while too small or too large charge clouds lead to a degradation of the resolution.
On the other hand, the network adapts implicitly to the individual charge cloud by extracting the features of the charge-sharing pattern of each cluster.
The absolute improvement is therefore largest at short propagation distances (Fig.~\ref{fig:resolution}(b)) with compact charge clouds, where the global mapping deviates most from the actual charge sharing.
Since the fraction of such photons increases with the photon energy, the advantage of the deep learning models is expected to become more pronounced at higher energies: in the 650~\textmu m-thick sensor, the fraction of the absorbed photons with a propagation distance below 250~\textmu m increases from about 11\% at 12~keV to about 30\% at 20~keV, while it is below 1\% at 8~keV.

\textbf{From simulation to measurement}: 
The models are trained exclusively on simulated events, and their agreement with the experimental data relies on the two prerequisites introduced in Section~\ref{sec:introduction}, namely a charge transport simulation including the repulsion effect \cite{XIE2026170894} and a pixel-wise gain and nonlinearity calibration \cite{XIE2026171227}.
The agreement between simulation and measurement is validated by the average two-photon clusters in Fig.~\ref{fig:2ph_average_sample}, which are consistent within 2\% in terms of the projected RMS.
The simulation thus facilitates the training by providing samples with the ground truth, which are difficult to obtain from measurements especially for pile-up events.

To compare simulation and measurement regarding the achievable resolution, the knife-edge measurements at both occupancies were also simulated at the frame level and analysed with the same reconstruction, ESF construction and fit as the measured data.
The results are summarized in Table~\ref{tab:sim-vs-exp}: the measured ESF widths are systematically larger than the simulated ones.
This difference is attributed to experimental contributions such as the quality of the knife edge, the residual angle between the beam and the sensor normal.
In particular, pixels carrying a signal exhibit additional dynamic noise beyond the pedestal fluctuation modelled in the simulation, which is expected to degrade the energy resolution and thus the position reconstruction.

Assuming that these contributions add a single, method-independent term $\sigma_{\mathrm{ext}}$ in quadrature, the expected ESF width is
\begin{equation}
  \sigma_{\mathrm{exp}}^{2} = \sigma_{\mathrm{sim}}^{2} + \sigma_{\mathrm{ext}}^{2}, 
  \label{eq:quadrature}
\end{equation}
where $\sigma_{\mathrm{sim}}$ is the width obtained from the simulated knife-edge measurement.
From the single-photon result of the eta interpolation at 0.74\% occupancy, $\sigma_{\mathrm{ext}} = 0.97 \pm 0.04$~\textmu m is obtained.
With this value, Eq.~(\ref{eq:quadrature}) reproduces the measured widths of all deep learning configurations within the uncertainties.
Since $\sigma_{\mathrm{ext}}$ is determined from the eta interpolation alone, this agreement shows that the resolution gain predicted by the simulation is fully retained in the measurement, which validates the simulation-based methodology.

\begin{figure}
  \centering
  \includegraphics[width=0.70\linewidth]{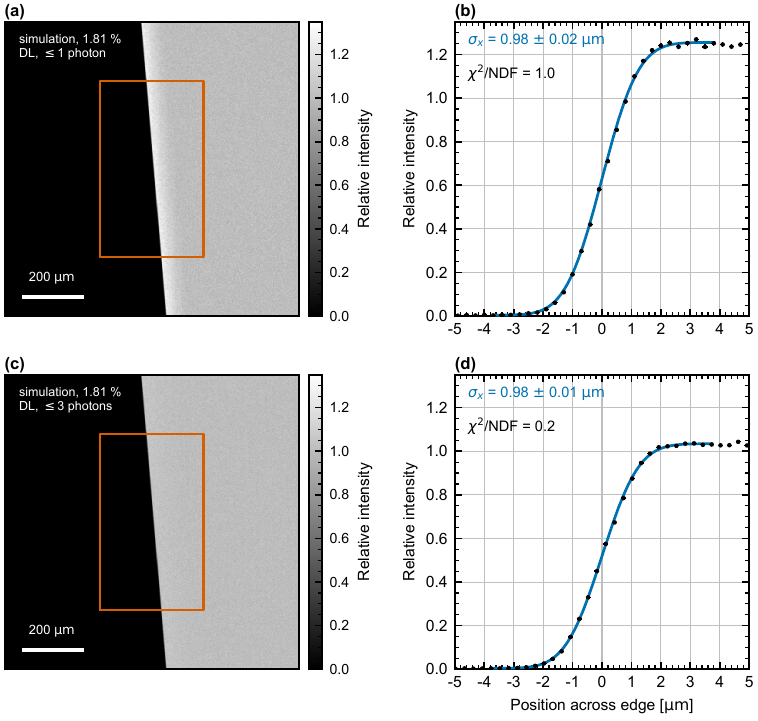}
  \caption{Simulated knife-edge measurement at 12~keV and 1.81\% occupancy, reconstructed with the deep learning models accepting only single-photon clusters (a) and clusters with up to three photons (c) with the region of interest indicated by the orange box.
  (b, d) plot the corresponding edge spread functions and error-function fits.
  The bright rim along the edge observed in Fig.~\ref{fig:exp-highflux}(a) is reproduced for single-photon acceptance and is removed when the pile-up clusters are reconstructed.}
  \label{fig:sim-knife-edge}
\end{figure}

\begin{table}[t]
  \centering
\caption{
  Measured ESF widths compared with the widths expected from Eq.~(\ref{eq:quadrature}) with $\sigma_{\mathrm{ext}} = 0.97$~\textmu m.
  The simulated widths are obtained from the frame-level simulation of the knife-edge measurements and include the 2.5~\textmu m binning, the non-Gaussian tails of the residual distributions and, for multi-photon acceptance, the mixture of multiplicities.
  For the 1.81\% occupancy, only the results with multi-photon acceptance are listed, since the single-photon results are affected by the bright rim discussed in Section~\ref{sec:exp-highflux} and the resulting worse ESF fitting quality.
}

  \label{tab:sim-vs-exp}
  \begin{tabular}{lccccc}
    \hline
    Occupancy & Method & \begin{tabular}[b]{@{}c@{}}Accepted\\ multiplicity\end{tabular} & \begin{tabular}[b]{@{}c@{}}Simulated\\ ESF width\end{tabular} & \begin{tabular}[b]{@{}c@{}}Expected\\ ESF width\end{tabular} & \begin{tabular}[b]{@{}c@{}}Measured\\ ESF width\end{tabular} \\
    \hline
    0.74\% & eta & $\leq$1 photon  & $1.19 \pm 0.03$~\textmu m & $1.54$~\textmu m$^{\dagger}$ & $1.54 \pm 0.02$~\textmu m \\
    0.74\% & DL  & $\leq$1 photon  & $0.97 \pm 0.02$~\textmu m & $1.37 \pm 0.04$~\textmu m & $1.36 \pm 0.02$~\textmu m \\
    0.74\% & DL  & $\leq$2 photons & $0.98 \pm 0.03$~\textmu m & $1.38 \pm 0.04$~\textmu m & $1.38 \pm 0.02$~\textmu m \\
    0.74\% & DL  & $\leq$3 photons & $0.97 \pm 0.02$~\textmu m & $1.37 \pm 0.03$~\textmu m & $1.38 \pm 0.02$~\textmu m \\
    \hline
    1.81\% & DL  & $\leq$2 photons & $0.98 \pm 0.01$~\textmu m & $1.38 \pm 0.03$~\textmu m & $1.37 \pm 0.02$~\textmu m \\
    1.81\% & DL  & $\leq$3 photons & $0.98 \pm 0.01$~\textmu m & $1.38 \pm 0.03$~\textmu m & $1.39 \pm 0.02$~\textmu m \\


    \hline
  \end{tabular}

  \vspace{1mm}
  {\footnotesize $^{\dagger}$Used to determine $\sigma_{\mathrm{ext}}$.}
\end{table}
\textbf{Necessity of pile-up reconstruction}: Even when only single-photon clusters are accepted, the number of reconstructed photons per pixel and frame is almost doubled by increasing the occupancy from 0.74\% to 1.81\%, which raises the question whether a pile-up reconstruction is necessary when the dose on the sample is not a limiting factor.
The rejection of pile-up clusters, however, makes the reconstruction efficiency depend on the local intensity, as photons in bright regions are more likely to form pile-up clusters than photons next to dark regions.
This effect produces the bright rim along the edge in Fig.~\ref{fig:exp-highflux}(a), with a relative intensity of 1.24 in the corresponding edge spread function.
This rim cannot be removed by the flat-field correction, since the flat-field measurement does not contain the intensity variations due to the edge.
The same effect is reproduced by the frame-level simulation, as shown in Fig.~\ref{fig:sim-knife-edge} for the 1.81\% occupancy.
With single-photon acceptance, the relative intensity next to the edge rises to 1.25, and it returns to 1.03 when clusters with up to three photons are accepted, in agreement with the measurement (1.24).
Such a nonlinear response distorts the image contrast and becomes more pronounced at higher occupancies.
Accepting clusters with up to three photons removes the rim and increases the number of reconstructed photons by a further factor of 1.55 at the same frame rate, which reduces both the measurement time and the data volume.

\textbf{Limitations of the method}: According to the curves of Fig.~\ref{fig:occupancy}, the three-photon hypothesis covers more than 93\% of the photons up to an occupancy of 1.81\%.
Above this occupancy, clusters with a higher multiplicity become relevant and additional models would be required, leading to the overhead of training and maintaining multiple models.
Moreover, the method cannot be pushed to arbitrarily high pile-up.
To contain all $N$ photons of a cluster, the area of the $3N \times 3N$ window grows quadratically with $N$, whereas the charge of each photon still covers only a few pixels, so that the number of pixels containing charge grows only linearly.
The fraction of pixels containing charge therefore decreases roughly as $1/N$, i.e.\ the input becomes increasingly sparse, and a larger window is also more likely to include pixels of neighbouring clusters.
Larger models are therefore required to process a sparser signal, at a rapidly decreasing inference throughput.
Such a categorization of the clusters furthermore relies on the monochromaticity of the beam, so it could not be used in experiments where the beam is polychromatic, such as full-field fluorescence imaging.

\textbf{Computational cost}: The inference throughputs of the deep learning models are 12~M events/s, 2.2~M events/s and 0.9~M events/s for the single-photon, two-photon and three-photon models, respectively, on an NVIDIA RTX 4090 GPU.
As already observed for the calibration \cite{XIE2026171227}, the total processing time of the offline analysis is dominated by reading the data from disk and by the cluster finding algorithm, and the additional cost of the deep learning models is therefore not significant.
An implementation of the integrated pipeline on GPU or FPGA is envisaged for online reconstruction.

\section{Conclusion and outlook}
\label{sec:conclusion}
Deep learning models have been developed to reconstruct the incident positions of X-ray photons in the M\"ONCH detector.
Simulations show that the deep learning models outperform the eta interpolation for single-photon events, with a residual RMS of 0.78~\textmu m compared with 1.07~\textmu m.
The improvement is understood as the ability of the network to adapt to the size of the individual charge cloud, while the eta interpolation forms a single mapping for all clusters with different absorption depths and charge cloud sizes.
Dedicated models resolve clusters containing two and three photons, with a residual RMS of 0.86~\textmu m and 1.09~\textmu m, respectively.
These results were validated with knife-edge measurements.
At an occupancy of 0.74\%, the width of the edge spread function is reduced from 1.54~\textmu m via the eta interpolation to 1.36~\textmu m via deep learning.
At an occupancy of 1.81\%, which is 2.4 times higher, the reconstruction of clusters containing up to three photons increases the reconstruction efficiency from 59\% to 93\% while the spatial resolution is preserved at 1.39~\textmu m (1.18~\textmu m without the binning contribution).
The usable photon flux is extended by a factor of 2.4 without noticeable compromise in reconstruction efficiency and spatial resolution, equivalent to an increase in the frame rate of the detector by the same factor.

For polychromatic cases, such as full-field fluorescence imaging \cite{10.1117/1.OE.56.4.043106}, the number of photons contained in a cluster can no longer be determined analytically from the cluster energy, which motivates a frame-based model that predicts the positions and the energies of the photons directly from the calibrated frames, bypassing the cluster finding step.
Such a model, together with the integration of the reconstruction into the data acquisition pipeline of the full-size M\"ONCH detector \cite{Heymes2025}, is the subject of ongoing work.
Finally, the methodology provides a tool to explore detector designs, where the achievable spatial resolution can be predicted and different design parameters can be optimized.


\section*{Declaration of competing interest}
The authors declare that they have no known competing financial interests or personal relationships that could have appeared to influence the work reported in this paper.

\section*{Acknowledgments}
We acknowledge MAX IV Laboratory for time on Beamline DanMAX under Proposal 20251632. Research conducted at MAX IV is supported by the Swedish Research council under contract 2018-07152, the Swedish Governmental Agency for Innovation Systems under contract 2018-04969, and Formas under contract 2019-02496. DanMAX is funded by the NUFI grant no. 4059-00009B.
We also thank Mads Ry J{\o}rgensen and Innokenty Kantor for their kind support during the beamtime.

\section*{Data availability}
The training and inference code, the trained model weights, simulation data used for parameterization and massive generation, and a subset of the measurement data are available at \url{https://doi.org/10.5281/zenodo.22807453}.

\bibliographystyle{elsarticle-num}

\bibliography{references}




\end{document}